\documentclass[sigconf, screen]{acmart}

\usepackage{float}
\usepackage[utf8]{inputenc} % allow utf-8 input
\usepackage[T1]{fontenc}    % use 8-bit T1 fonts
\usepackage{hyperref}       % hyperlinks

\usepackage{url}            % simple URL typesetting
\usepackage{booktabs}       % professional-quality tables
\usepackage{amsfonts}       % blackboard math symbols
\usepackage{nicefrac}       % compact symbols for 1/2, etc.
\usepackage{microtype}      % microtypography
\usepackage{tcolorbox}
\usepackage{adjustbox}
\usepackage{multirow}
\usepackage{graphicx}
\usepackage{subcaption}
\definecolor{bg}{HTML}{282828}
\usepackage{pmboxdraw}
\usepackage{algorithmic}
\usepackage{listings}
\usepackage[linesnumbered,ruled,vlined]{algorithm2e}
\def\BibTeX{{\rm B\kern-.05em{\sc i\kern-.025em b}\kern-.08em
    T\kern-.1667em\lower.7ex\hbox{E}\kern-.125emX}}

\usepackage{enumitem}
\usepackage{amsmath}
\usepackage[keeplastbox]{flushend}

\usepackage{natbib}
\setcopyright{acmcopyright}
\acmPrice{15.00}
\acmDOI{10.1145/3540250.3558959}
\acmYear{2022}
\copyrightyear{2022}
\acmSubmissionID{fse22ind-p60-p}
\acmISBN{978-1-4503-9413-0/22/11}
\acmConference[ESEC/FSE '22]{Proceedings of the 30th ACM Joint European Software Engineering Conference and Symposium on the Foundations of Software Engineering}{November 14--18, 2022}{Singapore, Singapore}
\acmBooktitle{Proceedings of the 30th ACM Joint European Software Engineering Conference and Symposium on the Foundations of Software Engineering (ESEC/FSE '22), November 14--18, 2022, Singapore, Singapore}

\begin{document}

\newcommand{\ie}{\textit{i.e.,}~}
\newcommand{\eg}{\textit{e.g.,}~}
\newcommand{\etc}{\textit{etc.}~}
\newcommand{\etal}{\textit{et al.}~}

%Comments
\newcommand{\nb}[2]{
    \fbox{\bfseries\sffamily\scriptsize#1}
    {\sf\small$\blacktriangleright$\textit{#2}$\blacktriangleleft$}
}

\newcommand\ANDREI[1]{\textcolor{red}{\nb{ANDREI}{#1}}}
\newcommand\MICHELE[1]{\textcolor{blue}{\nb{MICHELE}{#1}}}
\newcommand\COLIN[1]{\textcolor{green}{\nb{COLIN}{#1}}}

\newcommand{\approach}{{\sc AthenaTest}\xspace}
\newcommand{\dataset}{{\sc Methods2Test}\xspace}

%%% Coloring the comment as blue
\newcommand\mycommfont[1]{\footnotesize\ttfamily\textcolor{blue}{#1}}
\SetCommentSty{mycommfont}

\SetKwInput{KwInput}{Input}                % Set the Input
\SetKwInput{KwOutput}{Output}              % set the Output

% displaying code
\lstdefinestyle{myJavaStyle}{
  frame=tb,
  float=*,
  language=java,
  aboveskip=3mm,
  belowskip=3mm,
  showstringspaces=false,
  columns=flexible,
  basicstyle={\small\ttfamily},
  numbers=none,
  numberstyle=\tiny\color{gray},
  keywordstyle=\color{blue},
  commentstyle=\color{dkgreen},
  stringstyle=\color{mauve},
  frame=single,
  breaklines=true,
  breakatwhitespace=true,
  tabsize=3,
}

\title[Exploring and Evaluating Personalized Models for Code Generation]{Exploring and Evaluating Personalized Models\\for Code Generation}

% \author{A. Zlotchevski$^1$\thanks{Work done while interning at Microsoft.} , D. Drain$^2$\thanks{Work done while working at Microsoft.} , A. Svyatkovskiy$^3$, C. Clement$^3$, N. Sundaresan$^3$, M. Tufano$^3$  \\
% $^1$McGill University, $^2$Anthropic, $^3$Microsoft\\}
%\texttt{\{v-andreiz, dadrain, mitufano, neels\}@microsoft.com}}

\settopmatter{authorsperrow=3}

\author{Andrei Zlotchevski}
\affiliation{%
  \institution{McGill University}
  \city{Montreal}
  \state{Quebec}
  \country{Canada}
}
\email{andrei.zlotchevski@mail.mcgill.ca}

\author{Dawn Drain}
\affiliation{%
  \institution{Anthropic}
  \city{San Francisco}
  \state{CA}
  \country{USA}
}
%\authornote{Work done while working at Microsoft.}
\email{dawn@anthropic.com}
%\email{dawn.drain@microsoft.com}

\author{Alexey Svyatkovskiy}
\affiliation{%
  \institution{Microsoft}
  \city{Redmond}
  \state{WA}
  \country{USA}
}
\email{alsvyatk@microsoft.com}

\author{Colin Clement}
\affiliation{%
  \institution{Microsoft}
  \city{Redmond}
  \state{WA}
  \country{USA}
}
\email{coclemen@microsoft.com}

\author{Neel Sundaresan}
\affiliation{%
  \institution{Microsoft}
  \city{Redmond}
  \state{WA}
  \country{USA}
}
\email{neels@microsoft.com}

\author{Michele Tufano}
\affiliation{%
  \institution{Microsoft}
  \city{Redmond}
  \state{WA}
  \country{USA}
}

\email{mitufano@microsoft.com}
%\email{michele.tufano@microsoft.com}

%\renewcommand{\shortauthors}{}

\begin{abstract}

Large Transformer models achieved the state-of-the-art status for Natural Language Understanding tasks and are increasingly becoming the baseline model architecture for modeling source code. Transformers are usually pre-trained on large unsupervised corpora, learning token representations and transformations relevant to modeling generally available text, and are then fine-tuned on a particular downstream task of interest. While fine-tuning is a tried-and-true method for adapting a model to a new domain -- for example, question-answering on a given topic -- generalization remains an on-going challenge. In this paper, we explore and evaluate transformer model fine-tuning for personalization. In the context of generating unit tests for Java methods, we evaluate learning to personalize to a specific software project using several personalization techniques. We consider three key approaches: (i) \textit{custom} fine-tuning, which allows all the model parameters to be tuned; (ii) \textit{lightweight} fine-tuning, which freezes most of the model's parameters, allowing tuning of the token embeddings and softmax layer only or the final layer alone; (iii) \textit{prefix} tuning, which keeps model parameters frozen, but optimizes a small project-specific prefix vector. Each of these techniques offers a trade-off in total compute cost and predictive performance, which we evaluate by code and task-specific metrics, training time, and total computational operations. We compare these fine-tuning strategies for code generation and discuss the potential generalization and cost benefits of each in various deployment scenarios.

\end{abstract}

% \begin{IEEEkeywords}
% Code Generation, Deep Learning
% \end{IEEEkeywords}




\maketitle

%   <concept>
%       <concept_id>10011007.10011074.10011099.10011102.10011103</concept_id>
%       <concept_desc>Software and its engineering~Software testing and debugging</concept_desc>
%       <concept_significance>500</concept_significance>
%       </concept>
%   <concept>
%       <concept_id>10010147.10010178.10010179.10010180</concept_id>
%       <concept_desc>Computing methodologies~Machine translation</concept_desc>
%       <concept_significance>300</concept_significance>
%       </concept>
%  </ccs2012>

% \ccsdesc[500]{Software and its engineering~Software testing and debugging}
% \ccsdesc[300]{Computing methodologies~Machine translation}

% \keywords{software testing, unit test, neural networks}

\section{Introduction}

It is well-known that even the best models can fail to generalize properly to new domains, and even to new users of said models. For example, a model trained to answer questions in general may not answer StackOverflow questions as well as the questions in the training domain, or a software developer in an Enterprise environment with private code may have libraries and attribute name which differ from public source code used to train a code synthesis model. 

The current dominant paradigm in Natural Language Processing (NLP) modeling is to pre-train a large transformer model~\citep{DBLP:journals/corr/VaswaniSPUJGKP17} on a large corpus and then fine-tune it on a particular task of interest. For example, a question-answering (Q\&A) model is generally first pre-trained on a large corpus of textual data for the specific language (\eg Wikipedia, and news articles in English), then fine-tuned on a task-specific dataset of paired questions and corresponding answers. The pre-training process aims at learning semantic vector representation of the language and words, while the fine-tuning process specializes the model for a specific domain.

Transformer models are also increasingly the baseline architecture used for code generation tasks, such as writing methods from natural language description \citep{clement2020pymt5,austin2021program,chen2021evaluating}, or generating test cases from the focal method under test \citep{tufano2021unit}. Similarly for NLP tasks these models are pre-trained on a large corpus of natural text and publicly available source code and then fine-tuned on a specific code-related task. Further, these models also may not generalize to new domains of interest, and can benefit from task or even user-specific fine-tuning, here called customization or personalization. Customization is particularly relevant for code generation models since it provides several benefits:
\begin{itemize}
    \item allows fine-tuning on source code data that may not be available when training a base model (\eg private repositories or internal codebases), enabling improved overall performances on codebases with proprietary dependencies and code styles;
    \item the opportunity to improve data privacy by considering private or sensitive data only during the customization process on the client side;
    \item the opportunity to reduce deployment cost as customized models can offer better user performance without increasing model size.
\end{itemize}

Custom models can provide clear benefits to users and model providers. We envision serving tens or hundreds of thousands of custom models, but doing so presents several logistical hurdles, including the costs of training, storing, and loading these models into GPU memory for inference. Worse, memory costs will only be exacerbated when working with ever larger and more powerful models.

For these reasons, we investigate several customization approaches, some of which can dramatically reduce the memory footprint and amortized computational cost introduced by custom models. Specifically, we consider three fine-tuning approaches: (i) \textit{custom} fine-tuning, which allows all the model parameters to be tuned; (ii) \textit{lightweight} fine-tuning, which only optimizes the token embedding representations or the final softmax layer; (iii) \textit{prefix} tuning, which keeps language model parameters frozen, but optimizes a small project-specific vector prefix.

In our extensive empirical evaluation we found that all the customization strategies lead to significant model improvements on a target project in terms of both intrinsic and task-specific metrics. While there is no unambiguous winner among the customization strategies, each approach can provide specific benefits in particular deployment scenarios. This paper provides insights on these customization strategies, their benefits and drawbacks, as well as providing guidelines and suggestions on which one to use based on the training cost, memory and storage, number of users, and deployment scenarios.

% Holding training costs constant, our lightweight fine-tuning approach is able to capture X\% of the gains of end-to-end training, while reducing memory costs and I/O load by Y\%. When lightweight fine-tuning to convergence, which we observe takes longer than with end-to-end fine-tuning, we are able to capture X'\% of the validation loss and BLEU gains. [Would be very cool if we could say these approaches are complentary!!!]

% Finally, we investigate where the gains come from, and quantify the extent to which they stem from learning the custom repository's user-defined functions, variables, and classes, the available imports, and more stylistic matters such as casing conventions and the repository's typical testing approach.

% Motivations:
% \begin{itemize}
%     \item Improved Performances
%     \item Data Privacy
% \end{itemize}

% Challenges:
% \begin{itemize}
%     \item Training time
%     \item Training cost
%     \item Model storage
% \end{itemize}

% issues with the naive approach:
% imagine loading the model just for one-off inferencing lol

% imagine keeping all models loaded at a time

% storage costs are non-trivial

% an eye towards scaling

% cpu vs gpu memory cost vs. storage costs

\section{Motivation}
Software projects are often classified based on their architecture (\eg web, server, monolithic), domain (\eg finance, healthcare), topic or usages (\eg games, editors). In this context, several techniques have been proposed in the literature for the task of \textit{software categorization}, which aims at organizing projects into groups that broadly describe the behavior or topic of the software. MUDABlue \cite{kawaguchi2006mudablue}, relies on Latent Semantic Indexing (LSI), an Information Retrieval (IR) technique, to automatically categorize software systems in open source software repositories. For the same task, LACT \cite{tian2009using} uses Latent Dirichlet Allocation (LDA), and recently neural text classification with word embeddings has been used \cite{leclair2018adapting} to categorize similar software projects. 

While projects can be broadly categorized, each individual software project, apart from trivial forks and clones, have peculiar characteristics which make them unique. Codebases have different user-defined types, API usages, specific coding styles, and identifiers' preferences chosen by developers. These idiosyncrasies represent an additional level of complexity for models that aim at generating code for a variety of software projects. 

This is exacerbated by the fact that transformer models only receive a limited-size input during inference, often considering only the current source code file. This confined window of tokens (commonly set at 1024 tokens) cannot provide a complete view of the project with its peculiarities. An accurate generation requires information about packages, classes, APIs, and identifiers that are external to the portion of code provided as input. Thus, we argue for personalized models that can generate custom code for specific projects.

\begin{figure}[tbp]
    \centering
    \includegraphics[width=0.9\columnwidth]{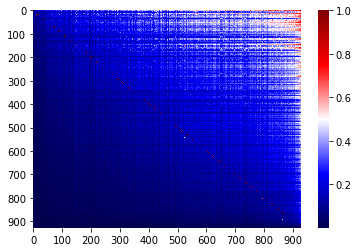}
    \caption{Heat-map displaying the ratios of shared tokens among software projects. Most projects share relatively few identifiers with other codebases.}
    \label{fig:tokens}
    \vspace{-0.5cm}
\end{figure}

As an exploratory study, we begin by observing projects' diversity in terms of tokens used in their source code. Specifically, we're interested in understanding the amount of shared tokens among different software projects. This could serve as an initial, rough, proxy metric to measure project diversity and potentially motivate the need for personalized models.

We select 930 Java software projects randomly sampled from GitHub, declaring an open source license, which have been updated within the last five years, and are not forks. These projects belong to the validation set of the open dataset Methods2Test \cite{tufano2021unit}. For each project, we collect all the available \texttt{.java} files combining them into a single-project corpus. Next, we remove licensing information and comments using regex, then tokenize the corpus using \texttt{gensim}~\cite{rehurek_lrec} tokenizer (with lowercase setting). From the list of tokens, we compute the set of unique tokens used within the project, and exclude the Java keywords from this set (similar to stopwords).

\begin{figure*}[htbp]
\vspace{-0.0cm}
\centering
\begin{subfigure}[b]{.2\linewidth}
\includegraphics[width=\linewidth]{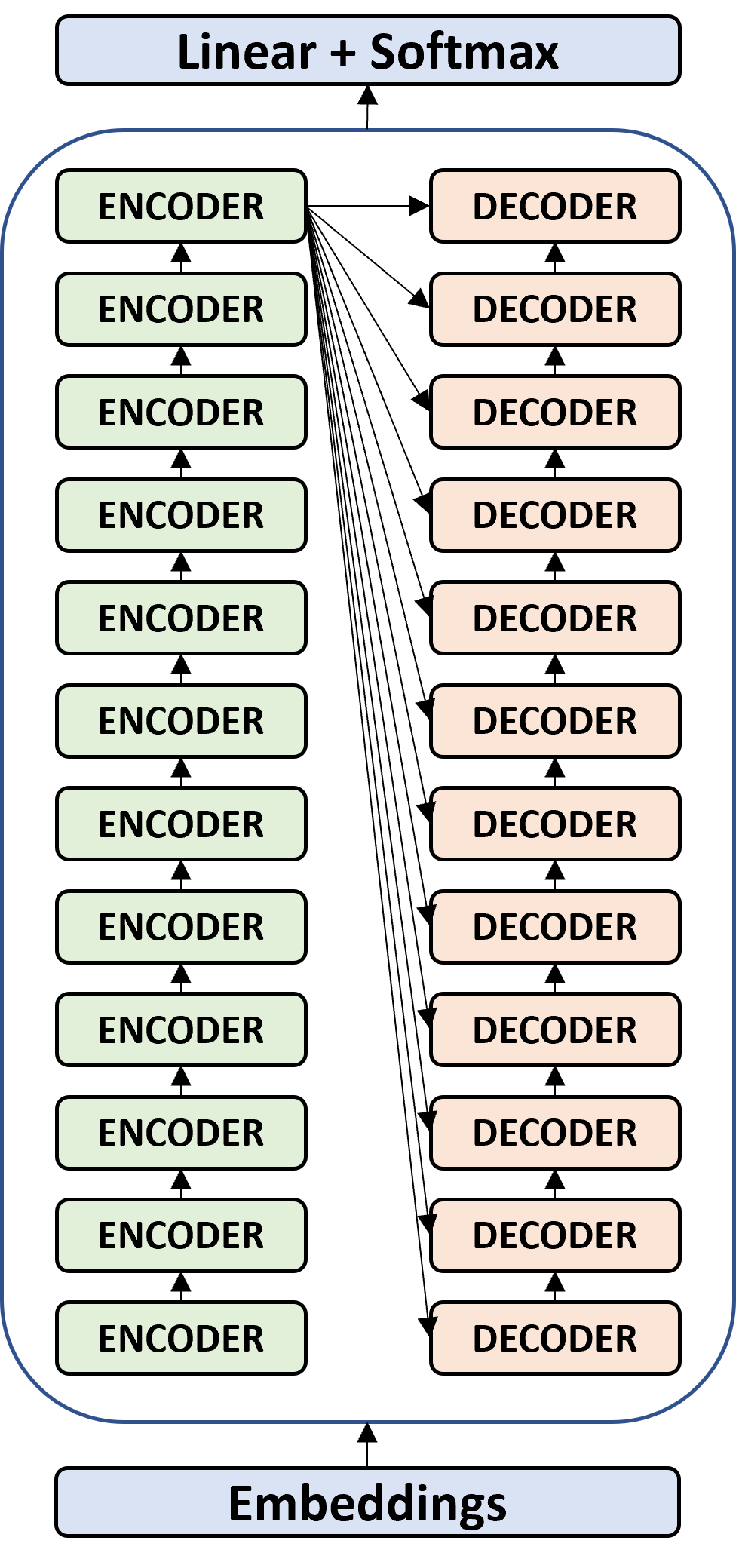}
\caption{Custom}\label{fig:CustomModel}
\end{subfigure}
\begin{subfigure}[b]{.2\linewidth}
\includegraphics[width=\linewidth]{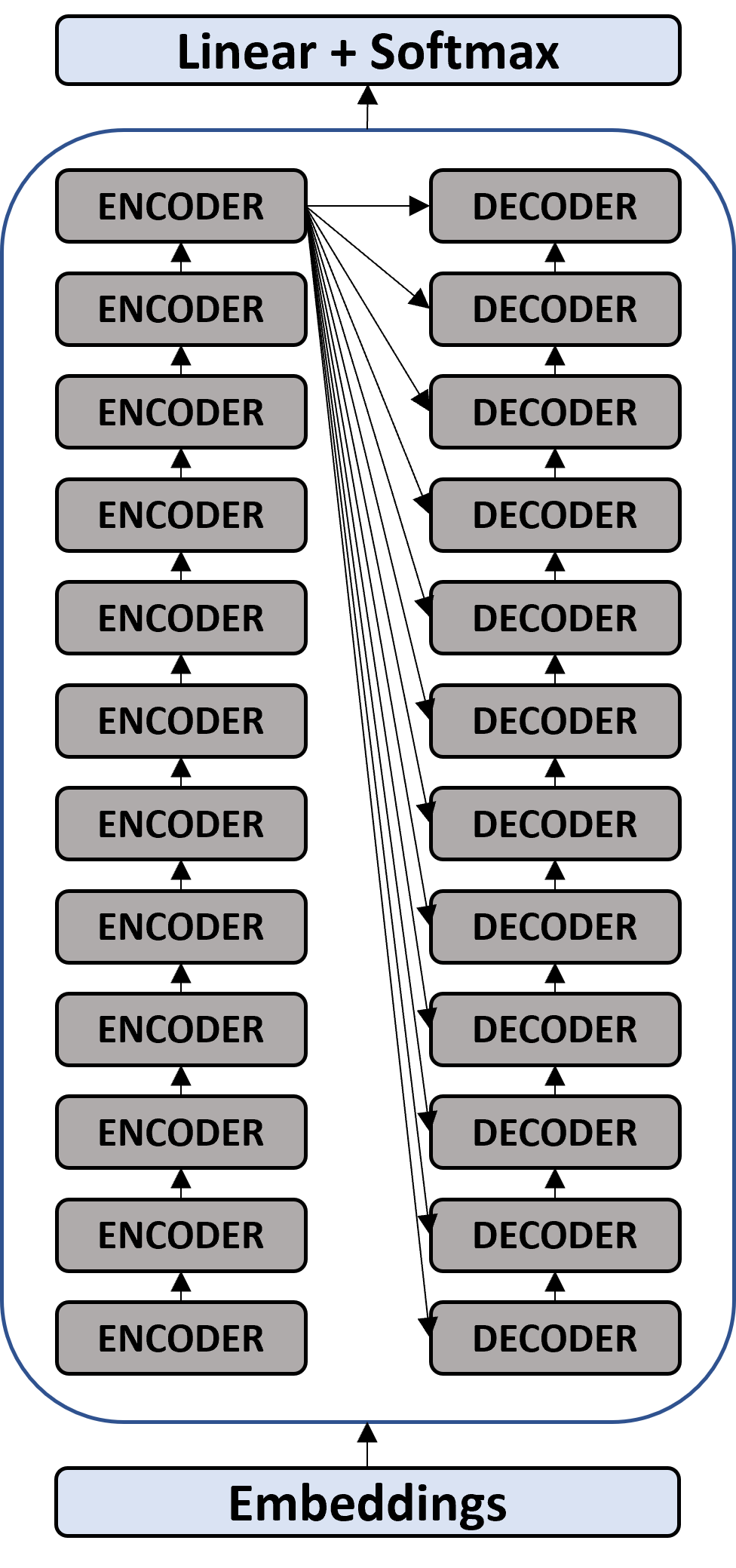}
\caption{L-EO}\label{fig:LightweightEmbedOutput}
\end{subfigure}
\begin{subfigure}[b]{.2\linewidth}
\includegraphics[width=\linewidth]{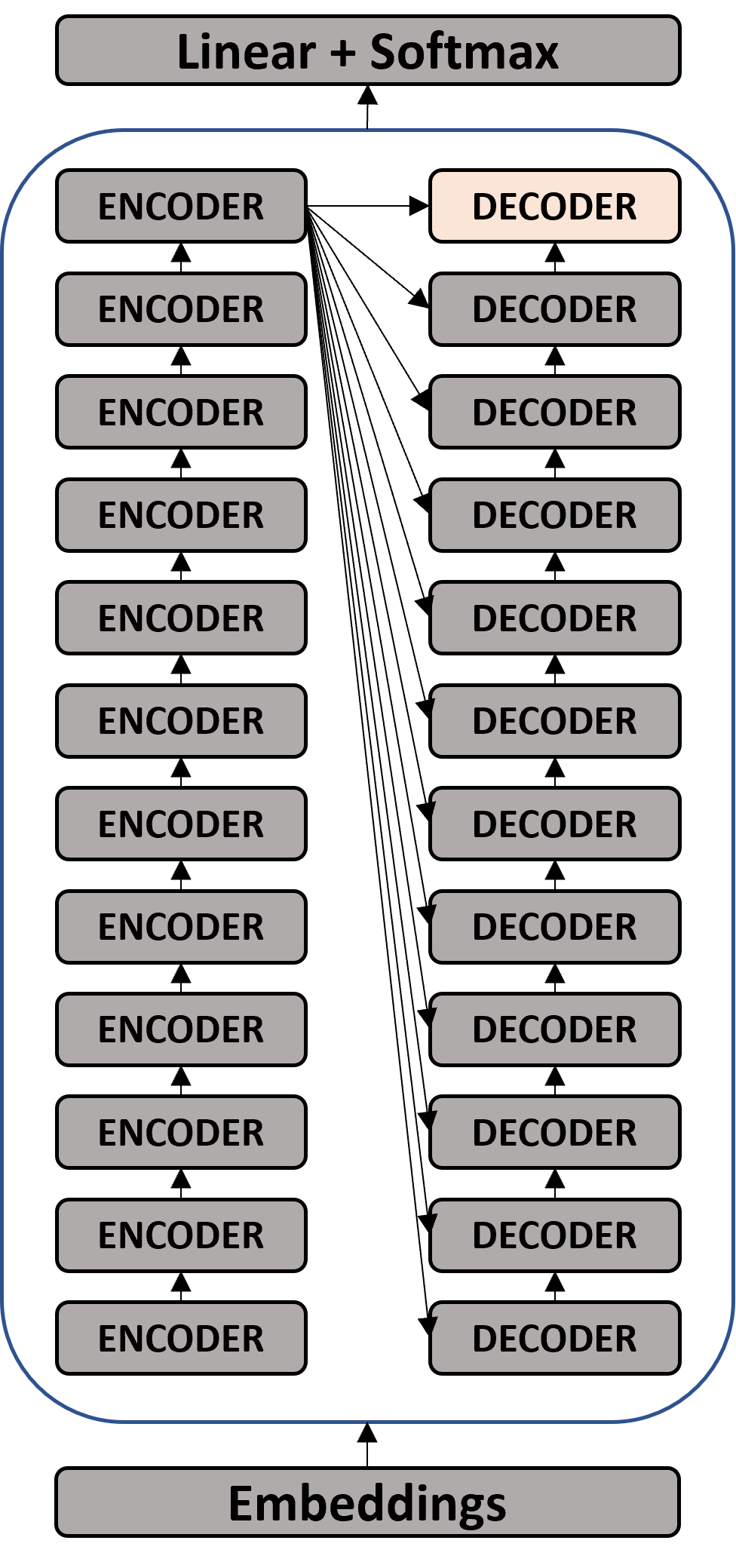}
\caption{L-LDB}\label{fig:LightweightDecoder}
\end{subfigure}
\begin{subfigure}[b]{.332 \linewidth}
\includegraphics[width=\linewidth]{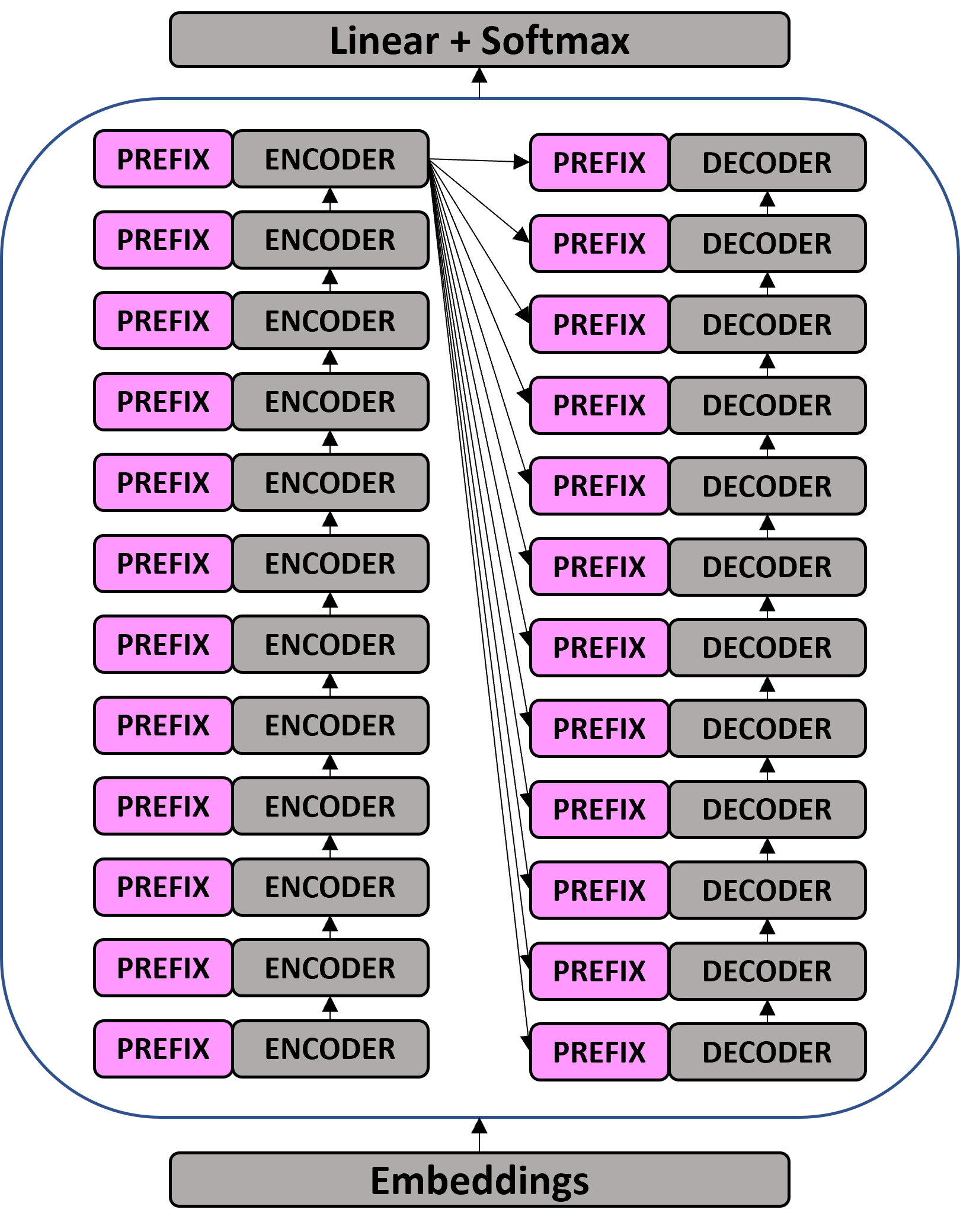}
\caption{Prefix}\label{fig:PrefixModel}
\end{subfigure}
\vspace{-0.0cm}
\caption{Overview of the Customization Approaches - Transformer models during fine-tuning, where the frozen parts of the model (not trainable) are displayed in gray: (a) Custom fine-tuning modifies all parameters during training; (b) L-EO trains only the embedding and output layers; (c) L-LDB allows to train only the parameters of the last decoder block; (d) Prefix tuning adds a trainable prefix to the encoder and decoder blocks.}
%\label{fig:ModelOverview}
\vspace{-0.0cm}
\end{figure*}

For each pair of projects $p_i$ and $p_j$, with token sets $T_i$ and $T_j$, we compute the shared token set $T_{i,j} = T_i \cap T_j$. Next, for both $p_i$ and $p_j$, we compute their corresponding ratios of shared tokens as follows: $R_{i,j} = |T_{i,j}|/|T_i|$ and $R_{j,i} = |T_{i,j}|/|T_j|$.

Figure \ref{fig:tokens} shows the ratio of shared tokens between each pair of projects as a heat-map. Projects are sorted in ascending order of the number of unique tokens used in their source code. Blue values indicate a low ratio of shared tokens (the darker, the lower), while red values indicate project pairs with substantial amount of shared tokens. The heat-map appears mostly blue, indicating that the majority of projects share relatively few tokens among each others. The upper-right corner shows pairs with higher ratios (white/red points), these are ratio computed for very small projects whose tokens are contained in very large projects, hence the corner position. Overall, the majority of project share relatively few identifiers with other projects. Specifically, if we consider all the values in the matrix $R$ except for the diagonal (\ie token shared with the project itself), a project on median shares only 13\% of its tokens with another project, and the third quartile of the distribution is below 23\%.

We consider this study only as a preliminary analysis into the diversity of software projects, which could motivate the need for personalized models. We acknowledge the limitations of this study, which could be extended considering different types of tokenizers, preprocessing steps, and metrics. In Sec. \ref{sec:experimental} we design an experimental study that analyzes in details the impact of personalization on the performances of transformer-based code generation models.

\section{Approach}
\label{sec:approach}
This section describes the proposed customization approach for code generation models. We begin by formally defining the customization process, then we provide details for each of the fine-tuning strategies.

\subsection{Customization Process}
 We use the term \textit{customization} to refer to the process of fine-tuning a model $m$, previously trained on a generic dataset for a task $t$, with the goal of improving its performance on a specific dataset $p$. The performance of a machine learning model $m$ on a dataset $p$ is measured by one or more evaluation functions $f(m,p)$, where $f$ can be either a maximization (\eg BLEU, top-k accuracy) or minimization (\eg perplexity) function. The customization process is designed to modify the trainable parameters of the model $m$, obtaining the model $m'$, such that the performance of $m'$ on $p$ is better than what was attained by $m$. Specifically, $f(m',p) > f(m,p)$ for maximization functions, or $f(m',p) < f(m,p)$ for minimization functions.
 
 In this work, $m$ is an encoder-decoder transformer model, $t$ is a code generation task, and $p$ is a target software project to which we intend to customize $m$.

\subsection{Custom Fine-tuning}
Custom fine-tuning is the most straightforward customization approach. The model to be customized is taken as is and trained on a selected project. All parameters are trainable during this process. Figure \ref{fig:CustomModel} shows the model during fine-tuning, where all the parameters from the encoder and decoder blocks, as well as embeddings and output layers can be modified.

\subsection{Lightweight Fine-tuning - Embeddings and Output Layer (L-EO)}
Fully fine-tuning a model for every project or user may be prohibitive in terms of storage and memory costs. As a result, we explore ways to mitigate these costs by reducing the number of parameters that vary from one custom model to another. In our lightweight fine-tuning experiments, we achieve this by freezing most parameters in the baseline model, and only keeping a small subset trainable. Figure \ref{fig:LightweightEmbedOutput} shows the Lightweight fine-tuning - Embeddings and Output Layer (L-EO) design, where most of the model parameters are frozen (displayed in gray), and we allow only the embedding and output layers parameters to be fine-tuned, following the approach in \cite{DBLP:journals/corr/abs-2103-05247}. 

\subsection{Lightweight Fine-tuning - Last Decoder Block (L-LDB)}
\begin{figure}[htbp]
    \centering
    \includegraphics[width=0.9\columnwidth]{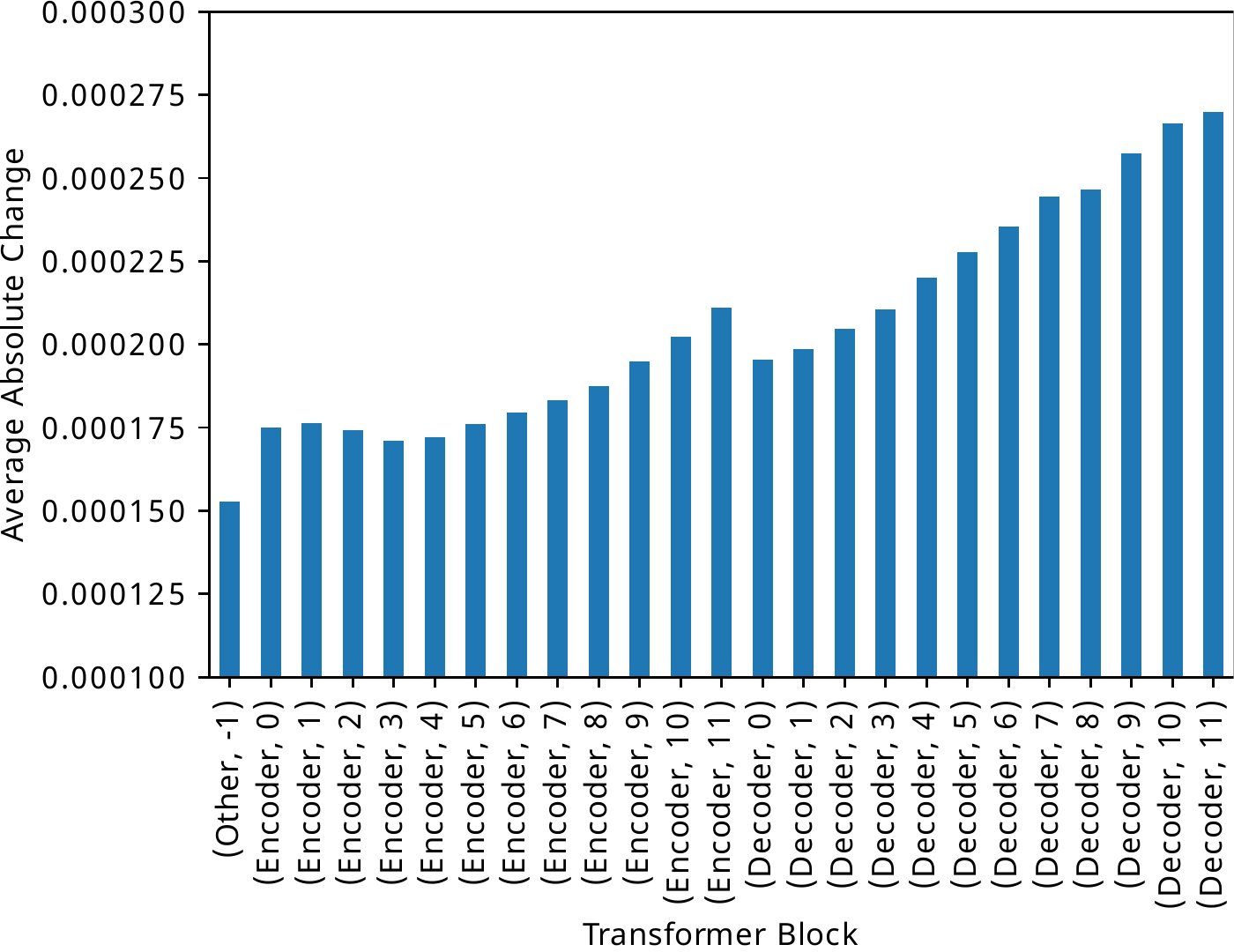}
    \caption{This figure shows the total average parameter change after fine-tuning to a new project domain, showing that the largest parameter changes occur in deeper parts of the model. This motivates our choice to try only fine-tuning the later layers of the model.}
    \label{fig:ParameterChangePerBlock}
    \vspace{-0.5cm}
\end{figure}

In this lightweight fine-tuning strategy, shown in Figure \ref{fig:LightweightDecoder} (L-LDB), most of the model's parameters are kept frozen, while only the parameters in the last decoder block are trainable, this includes: self-attention, encoder-decoder attention, layernorm and feedforward layers. This design decision of training only the last decoder block is motivated by experimental results analyzing the model's parameter changes during custom fine-tuning. Figure \ref{fig:ParameterChangePerBlock} reports the average absolute changes, during fine-tuning, in the parameters belonging to different Encoder and Decoder blocks for a BART model. We observe that, as we go through the transformer model, the average change in parameter values tends to increase, with the last decoder block showing the highest changes in parameter values. As a result, we hypothesize that it could be sufficient to tune the last decoder block and obtain performance improvements similar to the fully custom fine-tuned model.

% In this second lightweight fine-tuning experiment, the model is again frozen and only the parameters in the last decoder block are trainable: this includes the self-attention, encoder-decoder attention, layernorm and feedforward layers, for a total of 17M parameters. During the custom fine-tuning experiment, we examined which parameters are tuned the most. We observed that on average, as we go through the transformer model, the average change in parameter values tends to increase, as seen in Fig.\ref{fig:ParameterChangePerBlock}.
% As a result, we believe that it could be sufficient to tune the last decoder block and obtain performance improvements similar to the fully custom fine-tuned model. 

\subsection{Prefix Tuning}
Prefix tuning was first introduced by \citet{li-liang-2021-prefix}, with the goal of fine-tuning a general model for different tasks. The technique concatenates a sequence (prefix) of virtual tokens (trainable parameters) to the front of the input of every encoder and decoder block. In our context, the intuition behind this approach is that the prefix embeds the properties of a specific project, which allows the model to generate customized responses for that repository. Practically, we set the prefix length to 200 tokens, and thus with an embedding size of 1024, this gives a total of $1024\times 200 \times 24 \times 2 \approx 10$M trainable parameters. The prefix is initialized to the most frequent words in the repository for which the model is customized. % \MICHELE{We may need more details on this.}

\subsection{Trainable Parameters during Fine-tuning}
Table \ref{table:ApproachSummary} provides an overview of the number of total and trainable parameters involved in each customization process, in the case of a BART Transformer model with 406M parameters. Custom fine-tuning allows to train 100\% of the 406M available parameters in the model. During L-EO finetuning, instead, only 13\% (53M) parameters are trained. The L-LDB finetuning reduces the number of trainable parameters to 4.2\% (17M). Finally, Prefix tuning has the lowest number of trainable parameters, only 2.4\% (10M) of the total, but these are additional parameters added to the model (total 416M).

%, which reaches a total of 416M.

%\begin{table}[htbp]

\section{Experimental Design}
\label{sec:experimental}
The goal of our experimental design is to investigate whether custom models outperform the baseline model, leading to performance improvements in terms of intrinsic metrics (RQ$_1$), as well as extrinsic task-specific metrics (RQ$_2$). Next, we analyze and compare the different customization approaches in terms of training and compute costs (RQ$_3$) as well as model size and required storage for deployment.

In our case study, we chose Unit Test Case generation as our code generation task $t$, and \textit{AthenaTest} by \citet{tufano2021unit} as our baseline model $m$, which is a BART transformer model pre-trained on source code and English, and fine-tuned on Java unit test generation. The task is modeled as a translation task, where the input is a focal method (\ie method under test), and the output is a test case which tests the focal method's correctness. We randomly sample 20 projects from the test set, each of those representing the dataset $p$ on which a custom model is fine-tuned. Specifically, for each project $p$, we start from the baseline model $m$ and fine-tune four different custom models according to the four proposed fine-tuning strategies. For each project and fine-tuning strategy (\eg L-EO), we fine-tune and evaluate the models using 4-fold cross-validation. The models are trained until the best validation loss is reached, independently for every fold, every repository, and every customization approach. In total, we fine-tune and evaluate $20 (projects) \times 4 (approaches) \times 4 (folds) = 320$ models.

% In our case study, starting with one model $M$, we explore a total of four different customization processes $C$ on 20 different projects $P$. In this research project, we want to investigate if it is possible to obtain better performance by customizing a baseline model to a given repository, and if so, how large is the improvement (RQ1 and RQ2). While improved performance is the main goal of customizing models, it is important to note that these models are very costly to train, store, and deploy. We analyze the benefits of each customization approach with respect to these costs (RQ3 and RQ4).

% %% SUBSECTION HERE - Case study / model considered ...
% Our baseline model is \textit{AthenaTest} (cite Tufano). It is a BART model pre-trained on source code and English, and fine-tuned on Java unit test generation. For more details we refer to the original paper.

% For our experiments, we have sampled 20 repositories from the AthenaTest focal method/unit test dataset. We train the baseline model on the dataset within one repository. Since each dataset is relatively small,  we confirm the robustness of our results via 4-fold validation. The models are trained until the best validation loss is reached, independently for every fold, every repository, and every customization approach. In total, we investigate 4 different customization strategies, as described in Section \ref{sec:approach}.

\begin{table}[t]

\centering
\caption{Comparing the number of trainable parameters in each fine-tuning method.}
\vspace{-0.4cm}
\begin{tabular}{lll}
\toprule
\multirow{2}{*}{Customization Process} & \multicolumn{2}{c}{Parameters} \\ 
                & Total        & Trained  \\ \midrule
Custom          & 406M      & 406M (100\%)         \\ 
L-EO            & 406M      & 53M (13\%)          \\ 
L-LDB           & 406M      & 17M (4.2\%)        \\ 
Prefix          & 416M      & 10M  (2.4\%)      \\  \bottomrule
\end{tabular}
\label{table:ApproachSummary}
\vspace{-0.4cm}
\end{table}

\subsection{Dataset}
Table \ref{table:dataset} reports information about the 20 GitHub repositories sampled from the test set, which will be used to customize our models. The table shows (i) the Project ID, which will be used in the paper to reference a specific project; (ii) the project name; (iii) the project size in terms of disk usage; (iv) the popularity of the project in terms of number of stars obtained on GitHub; (v) and the dataset size, which corresponds to the number of data points for the unit test generation task (\ie pair of focal method and test case). The list of projects represent a diverse set of repositories with different size, domain, and popularity. They span from small personal projects (\eg \texttt{Tutorials} with 6 stars), to open source projects developed by large organizations such as Apache and Google.

%\ANDREI{Is the project size the total size of the files (in bytes) rather than the number of files?}

\begin{table*}[htbp]
\centering
\caption{Dataset - Projects used for customization}
\begin{tabular}{llrrr}
\toprule
Project ID   & Name                                & Project Size (MB)    &  Stars & Dataset Size\\
\midrule
26644682  & Talend Data Prep                    & 68.8     & 56    & 651 \\
40735368  & GeoTools                            & 62.4     & 8     & 653\\
107330274 & Titus Control Plane                 & 36.0     & 302   & 660\\
52972024  & Smart Actors                        & 57.8     & 22    & 704 \\
9714608   & Arakhn\^e Foundation Classes        & 17.9     & 13    & 753  \\
60701247  & Android Plugin for IntelliJ IDEA    & 1026.7   & 716   & 754 \\
14550159  & EverRest                            & 5.3      & 24    & 761 \\
9278888   & Brave                               & 18.8     & 2084  & 787 \\
66940520  & DHIS 2                              & 118.1    & 211   & 862 \\
33645537  & Tutorials                           & 34.4     & 6     & 878\\
62253355  & Mobi                                & 62.6     & 35    & 986 \\
155883728 & OakPAL                                & 15.0     & 9     & 1005 \\
4710920   & Apache Dubbo                        & 36.1     & 36231 & 1058 \\
29603649  & Wilma                               & 6.7      & 40    & 1074\\
42949039  & Herd                                & 227.2    & 127   & 1249 \\
1381673   & Drools                              & 176.7    & 3908  & 1394 \\
1244027   & ModeShape                           & 131.1    & 212   & 1550 \\
73948366  & AthenZ                              & 38.8     & 639   & 1920 \\
660443    & Chemistry Development Kit (CDK)     & 214.8    & 305   & 2591 \\
87849739  & Eclipse Ditto™ Project              & 52.5     & 311   & 2842 \\
\bottomrule
\end{tabular}
\vspace{0.2cm}
\label{table:dataset}
\vspace{-0.0cm}
\end{table*}

\subsection{\texorpdfstring{$\mathrm{RQ}_1$}{RQ1}: Intrinsic Evaluation Metrics}

\textbf{RQ$_1$: Do custom models obtain better performances on intrinsic metrics, such as BLEU and perplexity, w.r.t. the baseline?} To begin, we investigate how the different model customization approaches described in Sec. \ref{sec:approach} score on intrinsic metrics such as BLEU and perplexity. All approaches entail fine-tuning the baseline model to the dataset of a specific project, with the choice of parameters being tuned depending on the approach taken. The four variants are trained independently until the best validation loss is achieved. We report the BLEU4 score and the mean perplexity per token on the test fold, for all the 20 projects. Next, we perform statistical tests to investigate whether the observed differences between the baseline and custom models are significant, as well as differences among the customization approaches. Specifically, we rely on the Kruskal-Wallis test, a non-parametric statistical test.

\subsection{\texorpdfstring{RQ$_2$}{RQ2}: Task-specific Performances}
\textbf{RQ$_2$: Do custom models improve on performance metrics specific to unit test generation?}
We want to investigate how the different customization approaches compare with respect to the downstream task of generating unit tests. Beyond BLEU score and perplexity, we would like to see if custom models can produce the correct target code, how closely their unit tests mimic the repository style, or even if they can perfectly match the desired output.
\begin{itemize}
    \item \textit{Perfect Matches}: We compare the model's output string with the target developer-written unit test. If the two strings are identical, this is considered a perfect match. We do not take spacing and indentation into account as we are using a Java dataset (where indentation is not required). We report the proportion of perfect matches among the top 5 model predictions.
    \item \textit{Abstracted Code Matches}: We pass the model output and target output through the src2abs tool \citep{src2abs}, to obtain an abstracted version, masking variable names, method names, etc. We also do not distinguish between different objects of the same type.
    \item \textit{Coding Style}: For each project's custom model, we would like to determine how closely the model learns the developer's personal programming style and preferences. To this end, we extract the collection of all identifiers (\ie variables and functions' names) from the unit tests written by the developer as well as those generated by the models. We then pass these text outputs through a tf-idf vectorizer and compute the cosine similarity between them. This allows us to compare the developer's and the models' word usage. We examine the similarity between the developer's unit tests and the baseline and custom models generated tests. This scores the vocabulary similarity of the unit tests with the model generated code. 
\end{itemize}

    %\MICHELE{Andrei, please, double check my revision here. I believe we only show the similarity between the identifiers in dev tests and model tests, not the production code. Correct?} \ANDREI{What I meant is that we also consider the production code when weighing the words, but you are correct that we are not showing similarity of unit tests to it. Let me know if we should redo the tfidf scores. We could also mention that if we split words by CamelCase, then the baseline's score improves a lot, which is evidence that the baseline knows the words, but just not how to combine them into meaningful names. I don't have that graph currently.}

%  ORIGINAL VERSION
% \item \textit{Coding Style}: For each project's custom model, we would like to determine how closely the model learns the developer's personal programming style and preferences. To this end, we extract the collection of all identifiers from 1) the production corpus, 2) the unit tests written by the developer and 3) the model output. We then pass these text outputs through a tf-idf vectorizer and compute the cosine similarity between them. This allows us to compare the developer's and the models' word usage. We examine the similarity between the developer's unit tests and the production corpus, and contrast it with the model's output similarity to the production corpus. We also score the vocabulary similarity of the unit tests with the model generated code.

\bigskip

\begin{figure*}[htbp]
\centering
\begin{subfigure}[b]{.45\linewidth}
\includegraphics[width=\linewidth]{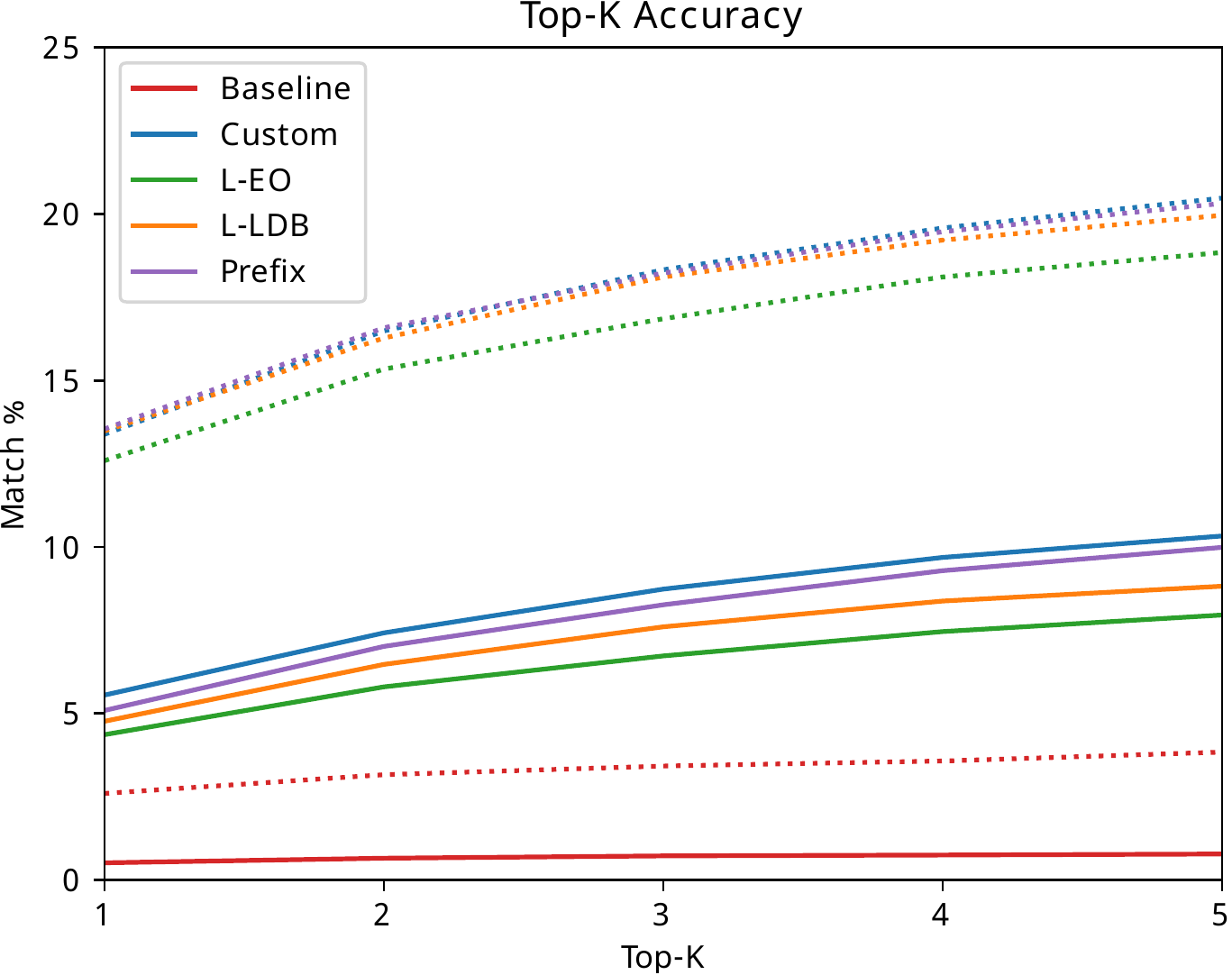}
\caption{Exact (solid) and abstracted (dotted) match rate}\label{plot:PerfectMatch}
\end{subfigure}
\begin{subfigure}[b]{.48\linewidth}
\includegraphics[width=\linewidth]{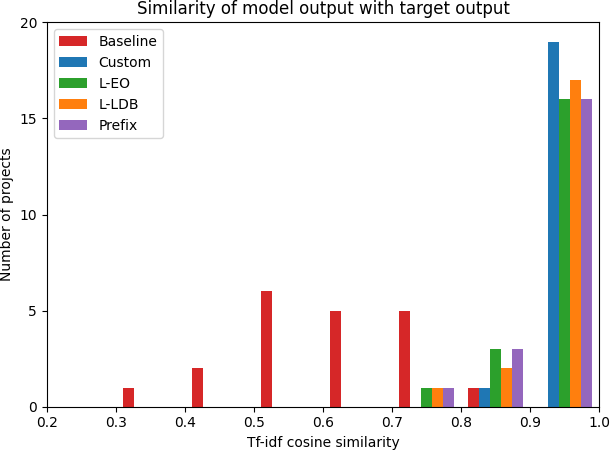}
\caption{Coding style as identifier similarity}\label{plot:coding_style}
\end{subfigure}
\caption{Task-specific metrics (a) custom models outperform the baseline in terms of perfect matches (solid line) and abstract matches (dotted line); (b) custom models generate code that uses identifiers (\ie variable and function names) that are more similar to the project codebase.}
\label{fig:task_perf}
\end{figure*}

\subsection{\texorpdfstring{RQ$_3$}{RQ3}: Training Cost Comparison} \label{subsection:compute}
\textbf{RQ$_3$: Given the same amount of compute, which custom models achieve the biggest performance improvement?} Since our four training regimes tune a different number of parameters, simply comparing the training time or number of optimization steps to reach the best validation loss may not be appropriate. For a model with $N$ parameters, we approximate the computation cost of a forward pass to be $C\approx2N$ floating point operations per training token, with an additional correction for embedding layers. The backward pass takes roughly twice the amount of compute, but it is unnecessary for layers that are frozen. For additional details, we refer to Table 1 in \cite{DBLP:journals/corr/abs-2001-08361}. We report the resulting compute in petaFLOPS-seconds.

%\ANDREI{Does this apply to us?}
%\ANDREI{Please help double check this!}

% --------- REMOVING THIS RQ4 - TALK ABOUT THIS IN DISCUSSION SECTION

% \subsection{RQ$_4$: Model size and storage}
% \textbf{RQ$_4$: How do the custom models compare with respect to memory requirements, both for storage as well as during inference?} In this research question we compare and discuss the memory requirements of the different customization strategies. We consider the memory storage costs, for the trainable parameters for each custom strategy (\ie parameters that are not frozen, thus different from the baseline model). Additionally, we discuss the differences also in memory requirements during inference, needed when loading a model on a GPU. \MICHELE{Should we remove this RQ and simply discuss this in the discussion section?}

% The lightweight fine-tuning and prefix tuning approaches freeze most (or all) of the original model's parameters. As a result, the storage costs of these custom models are greatly diminished, as only a fraction of the parameters need to be stored. We will do a more detailed calculation to obtain the cost savings.

% Memory comparison on storage and deployment of these three strategies.

\section{Results}

\subsection{\texorpdfstring{RQ$_1$}{RQ1}: Intrinsic Evaluation Metrics}
Table \ref{table:intrinsic_metrics} presents the results of custom models in terms of the intrinsic metrics: BLEU and perplexity. Specifically, for each project, we report the average BLEU and perplexity over the four folds, achieved on the test set by the different customization strategy, as well as the baseline model. We observe notable improvements in both metrics for every project w.r.t. the baseline, with BLEU going from 16.1 achieved by the baseline model to 36-38 by custom models. 

The statistical tests reported in Table \ref{table:statistical_test} confirm that the improvement observed by the four customization techniques are statistically significant w.r.t. the baseline ($p<1$e-7). However, we do not observe statistical significance in the differences among the customization strategies, meaning that, in terms of intrinsic metrics performances, the differences are within margin of error.

\begin{table*}[htbp]
\centering
\caption{The BLEU score and perplexity for the customization methods evaluated on the 20 projects in our test set.}
\small
\begin{tabular}{llllll|lllll}
\toprule
\multirow{2}{*}{Project} & \multicolumn{5}{c}{BLEU4} & \multicolumn{5}{c}{Perplexity} \\
          & Base      & Cust.  & L-EO    & L-LDB   & Prefix & Base & Cust.  & L-EO    & L-LDB   & Prefix \\ \midrule
          26644682 & 14.1 &   32.9 &   31.6 &   31.9 &    \textbf{34.0} & 1.275 & 1.212 & 1.208 & 1.238 &  \textbf{1.197} \\
          40735368 & 18.5 &   \textbf{30.7} &   29.0 &   29.4 &    29.1 & 1.276 & \textbf{1.186} & 1.197 & \textbf{1.186} &  1.194 \\
          107330274 & 14.8 &   \textbf{38.0} &   35.0 &   35.9 &    35.7 & 1.273 & \textbf{1.160} & 1.168 & 1.164 &  1.175 \\
          52972024 & 10.2 &   31.8 &   \textbf{33.2} &   32.2 &    30.1 & 1.271 & 1.142 & 1.146 & \textbf{1.135} &  \textbf{1.135} \\
          9714608 & 14.7 &   \textbf{41.0} &   38.1 &   40.4 &    40.2 & 1.263 & 1.155 & 1.145 & 1.150 &  \textbf{1.138} \\
          60701247 & 10.8 &   \textbf{28.9} &   24.4 &   25.9 &    26.6 & 1.267 & 1.187 & 1.190 & \textbf{1.172} &  1.176 \\
          14550159 & 20.0 &   \textbf{49.5} &   47.2 &   46.6 &    46.4 & 1.245 & 1.121 & 1.122 & \textbf{1.116} &  1.124 \\
          9278888 & 17.3 &   46.8 &   44.5 &   47.2 &    \textbf{47.8} & 1.272 & \textbf{1.137} & 1.152 & 1.138 &  1.140 \\
          66940520 & 17.4 &   \textbf{37.9} &   33.9 &   35.5 &    37.7 & 1.264 & 1.154 & 1.163 & 1.154 &  \textbf{1.150} \\
          33645537 & 17.0 &   30.4 &   31.2 &   \textbf{32.0} &    31.0 & 1.264 & 1.231 & 1.200 & \textbf{1.192} &  1.211 \\
          62253355 & 14.7 &   \textbf{48.0} &   45.7 &   47.3 &    \textbf{48.0} & 1.292 & \textbf{1.113} & 1.114 & 1.114 &  1.116 \\
          155883728 & 13.7 &   \textbf{41.3} &   37.5 &   39.3 &    39.5 & 1.238 & \textbf{1.132} & 1.148 & 1.146 &  1.140 \\
          4710920 & 28.2 &   \textbf{39.1} &   38.1 &   38.8 &    38.6 & 1.218 & 1.161 & 1.162 & 1.167 &  \textbf{1.160} \\
          29603649 & 19.1 &   \textbf{58.4} &   54.9 &   56.6 &    56.8 & 1.266 & \textbf{1.096} & 1.110 & 1.099 &  1.098 \\
          42949039 & 17.0 &   \textbf{38.2} &   37.7 &   37.5 &    37.3 & 1.238 & 1.154 & 1.152 & 1.154 &  \textbf{1.148} \\
          1381673 & 14.3 &   \textbf{33.3} &   29.3 &   30.9 &    30.8 & 1.261 & \textbf{1.133} & 1.152 & 1.138 &  1.138 \\
          1244027 & 19.6 &   \textbf{30.1} &   29.7 &   30.0 &    30.0 & 1.244 & \textbf{1.142} & 1.160 & \textbf{1.142} &  1.150 \\
          73948366 & 12.0 &   33.1 &   31.8 &   \textbf{34.0} &    33.6 & 1.267 & 1.161 & \textbf{1.157} & 1.159 &  1.164 \\
          660443 & 15.0 &   34.0 &   \textbf{37.2} &   36.5 &    34.3 & 1.281 & 1.180 & 1.170 & \textbf{1.169} &  1.177 \\
          87849739 & 13.4 &   45.1 &   47.0 &   \textbf{48.9} &    46.8 & 1.259 & 1.138 & 1.136 & \textbf{1.124} &  1.144 \\  \midrule
  Average & 16.1 &   \textbf{38.4} &   36.9 &   37.8 &    37.7 & 1.262 & \textbf{1.153} & 1.158 & \textbf{1.153} &  1.154 \\  \bottomrule
\end{tabular}
\label{table:intrinsic_metrics}
\end{table*}

\begin{table*}[htbp]
\centering
\vspace{0.5cm}
\caption{Kruskal-Wallis Test p-values testing the significance of the pairwise hypothesis that one customization method is superior than another. Custom strategies are significantly better than baseline.}
\small
\begin{tabular}{llllll|lllll}
\toprule
\multirow{2}{*}{} & \multicolumn{5}{c}{BLEU4} & \multicolumn{5}{c}{Perplexity} \\
          & Base      & Cust.  & L-EO    & L-LDB   & Prefix & Base & Cust.  & L-EO    & L-LDB   & Prefix \\ \midrule
Base &  - &  \textbf{3e-08} &  \textbf{3e-08} &  \textbf{3e-08} &  \textbf{3e-08} &  - &  \textbf{3e-08} &  \textbf{3e-08} &  \textbf{3e-08} &  \textbf{3e-08}\\
Cust. &  \textbf{3e-08} &  - &  0.4 &  0.7 &  0.7 &  \textbf{3e-08} &  - &  0.5 &  0.9 &  0.9\\
EO &  \textbf{3e-08} &  0.4 &  - &  0.5 &  0.7 &  \textbf{3e-08} &  0.5 &  - &  0.5 &  0.5\\
LDB &  \textbf{3e-08} &  0.7 &  0.5 &  - &  0.9 &  \textbf{3e-08} &  0.9 &  0.5 &  - &  0.8\\
Prefix &  \textbf{3e-08} &  0.7 &  0.7 &  0.9 &  - &  \textbf{3e-08} &  0.9 &  0.5 &  0.8 &  -\\
\bottomrule
\end{tabular}

\label{table:statistical_test}

\end{table*}

% COMMENT
% \COLIN{It's sort of trivial that all are better than the untrained one, maybe we should just compare to the baseline. If the p-value here is ~0.5 does that mean it's inconclusive? So the EO model is not conclusively better than the others?} \MICHELE{"mabe we should just compare to the baseline": we are indeed comparing to the baseline (non-customized). The point that we want to show is that all the customization approaches produce custom models that are statistically significantly better than the baselines, but there is no sign. diff. among each other (which is not a bad thing, it just means that one should choose the approach that makes more sense in that scenario/deployment)}

\subsection{\texorpdfstring{RQ$_2$}{RQ2}: Task-specific Performances}
The results in terms of task-specific performances are presented in Figure \ref{fig:task_perf}.
The plot \ref{plot:PerfectMatch} shows the top-k accuracy for perfect matches (solid line) and abstracted matches (dotted line), aggregated over the 20 projects. The baseline model outputs the same code structure (abstracted) in roughly 3\% of all cases, and virtually never produces the exact target output (<1\%). Moreover, its performance does not improve as we consider more predictions. All customization processes show significant improvement compared to the baseline. Specifically, these improvements are observed for every single project (full results will be available on our online appendix). Customized models produce the correct code structure as their top prediction in $\sim$13-14\% of instances, and a perfect match in $\sim$4-6\% of cases. They also tend to improve as we consider their top 5 predictions. Between the different customization processes, Custom consistently performs the best, closely followed by Prefix and L-LDB. When considering abstracted code matches, these three approaches are nearly identical. L-EO, however, performs slightly worse than the others.

Plot \ref{plot:coding_style} shows the distribution of tf-idf cosine similarity computed between identifiers used in the developers' written tests and the models' generated outputs. We observe that the distribution for custom models is skewed towards the higher values of cosine similarity. This result demonstrates that custom models tend to use variable and function names that are more similar to what developers used in their own tests.

%\caption{Coding style comparing hypothesis identifiers to true identifiers.}\label{plot:coding_style}

% \begin{figure*}[htbp]
\begin{figure*}[h] 
\centering
\includegraphics[width=0.8\linewidth]{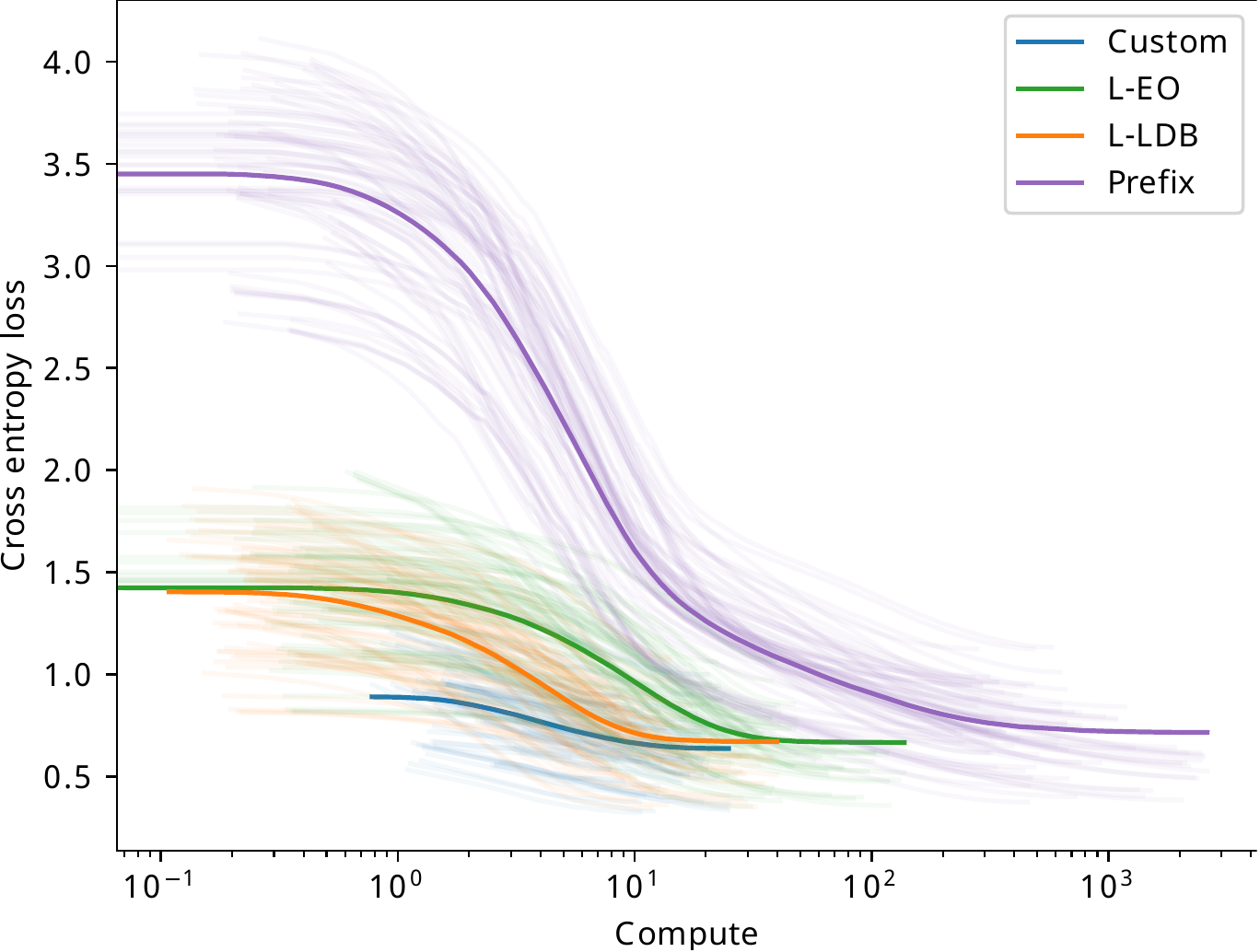}
\caption{Validation Loss vs Compute (PF-seconds) -  Light lines represent the validation loss curve for each individual project and fold, while the bold line represents the average for each custom strategy. Custom is the most efficient, lightweight approaches require slightly more compute to reach a comparable validation loss, while prefix is the least efficient, suffering from poor initialization.}\label{plot:BestLossCompute}
\end{figure*}

\subsection{\texorpdfstring{RQ$_3$}{RQ3}: Training Cost Comparison}
%We observe that in all experiments, the time required to achieve the best validation loss is highly correlated to the size of the dataset. Intuitively, given more data about a specific project or developer, the model can be more finely customized, and therefore naturally takes a longer amount of time to train. In fact, the correlation between the number of training steps and the number of test cases in the project is $0.87$, for both custom fine-tune and lightweight-fine-tune. %On a virtual machine with one GPU, each training step requires $0.68\pm 0.03$ seconds for custom fine-tune, and $0.51 \pm 0.02$ seconds for lightweight-fine-tune, which is about 25\% faster. However, lightweight-fine-tune converges to the best validation loss much more slowly than custom fine-tune.  %taking on average $15$ times more steps to achieve the best validation loss. 

For each customization process, we plot validation loss as a function of compute, as defined in section \ref{subsection:compute}. The results are presented in Figure \ref{plot:BestLossCompute}, where the light lines represent the validation loss curve for each individual project and fold, while the bold line represents the average for each custom strategy. First note that Custom achieves very large gains during the first epoch, as evidenced by the fact that its validation loss starts much lower than L-EO and L-LDB. Custom also outperforms other customization processes when given a limited amount of compute. However, we observe that beyond a certain amount of compute, Custom and L-LDB tend to achieve similar performances. In contrast, L-EO starts at the same validation loss as L-LDB but converges much slower to the best loss, requiring 2-3 times as much compute. 

Since the prefix parameters suffer from poor initialization, Prefix is the most expensive customization process. To overcome this problem, it is possible to first train the prefix on a large generic dataset. Then, given proper hyperparameter tuning, it is possible to substantially cut down compute cost for customizing the prefix.

\section{Discussion \& Lessons Learned}
The four customization strategies considered in this work are effective in improving a code generation model's performances on a given software project. Specifically, all custom models significantly outperform the baseline in terms of intrinsic metrics (\ie BLEU and perplexity) as well as task-specific metrics (\ie abstract and raw matches). While the differences among the customization approaches are not significant (no clear winner), each strategy offers specific advantages in different circumstances and deployment scenarios.

\textbf{\textit{Custom}} fine-tuning achieves the overall best performances and the customization process is relatively fast and efficient. This is somewhat expected, since this customization strategy allows all the model's parameters to be tuned on the specific project. This characteristic also leads to the major disadvantage of this approach: each custom model is an entire copy of the original model. Storage and inference costs could become prohibitive when serving many users with personalized custom models.

\textbf{\textit{Lightweight}} fine-tuning achieves good results while training fewer parameters. This allows to serve potentially many users with custom models which can be stored and loaded efficiently. Specifically, L-LDB trains fewer parameters than L-EO, however the latter could allow to deploy the embedding and output layers on the user side, with a privacy-preserving focus.

\textbf{\textit{Prefix}} fine-tuning trains the lowest number of parameters (only 2.4\% for a BART model), while improving over the baseline. However, it increases the total number of parameters of the model (prefixes are additional virtual tokens) and requires more compute time to achieve good performances, mostly due to the prefix initialization problem. On the bright side, this strategy allows to batch together requests from different users (with different prefixes), which can be processed by a single model, generating personalized outputs.

%\MICHELE{Should we have a table summarizing pro/cons?}

%\section{Deployment}
%\MICHELE{@ALL, I was thinking we may be able to add a diagram with the hybrid client/server deployment, to further motivate this work. However, I don't think it's possible to do so in the available space. We would have to describe the hybrid deployment for every custom strategy, and we don't have space. Also, if we just add a diagram, for one case, with little explanation of how it works, we may attract negative feedback/questions from reviewers. We will describe this deployment in the patent, this will help us lay down all the details, and could be used in the future for a follow-up paper (or extension). Still, I do believe we should try to give some intuition of different deployments we could have with the custom models. Any suggestions? Please feel free to add discussion on this!}

\section{Threats to Validity}

The major threats to our study relate to \textit{external validity}, which concerns the generalization of our findings. Our study is performed on a specific coding task (\ie test case generation) and for a single programming language (\ie Java). The findings could not generalize to all the coding tasks and different programming languages. However, our extensive empirical analysis investigating different personalization strategies in terms of several performance metrics, can provide guidelines for applying these techniques on different coding tasks, languages, and datasets. It is important to note that, while each coding task has its peculiarities, test case generation task involves the generation of complete methods, variable assignments, method calls, and different types of statements, thus could serve as a good generic coding task. Java language is among the most popular and similar to other programming languages such as C\# and Ruby. As part of our future works we intend to apply these personalization techniques to different coding tasks and languages.

\section{Related Work}

This work is related to two areas of the existing literature: neural source code generation and model personalization. Neural code generation has generated an intense recent interest in NLP, using Transformer models~\cite{vaswani2017attention} in particular for code completion~\cite{gptc,svyatkovskiy2019pythia,clement2020pymt5,raychev2014code,bruch2009learning,brockschmidt2018generative}, code synthesis from examples~\cite{chen2018towards}, natural language to code~\cite{clement2020pymt5,chen2018towards,austin2021program}, code feature summarization~\cite{hgnn_iclr21,moreno2013automatic,scalabrino2017automatically,wan2018improving,alon2018code2seq,moreno2014automatic}, code search~\cite{husain2019codesearchnet,feng2020codebert}, unit test generation~\cite{tufano2021unit} and even bug fixing~\cite{deepdebug_java} and detection~\cite{zhai2020cpc}. This paper naturally is an extension and evaluation of personalized unit test generations as studied by \citet{tufano2021unit}, and an important contribution to the understanding optimization in a deployment scenario.

Much of the previous literature on personalized models focuses on client-side training to keep data on device~\cite{shor2019personalizing,popov2018distributed}, and most work is in the domain of search query completion~\cite{jaech2018personalized}, natural language completion~\cite{popov2018distributed}, or even automated speech recognition~\cite{shor2019personalizing}. Naturally this work extends the domain of evaluation beyond natural language tasks and into the software engineering domain. This paper does not evaluate methods for client side training with restricted resources, however, as the most powerful large language models which enable state of the art code synthesis have 10-100 million parameters. At the time of writing such large models cannot be executed in a reasonable amount of time on most consumer laptops. We leave to future work extending these studies to models which have been pruned, quantized, distilled, and optimized to be ran in limited resource environments.

%We cam also discuss this (\url{https://ai.googleblog.com/2020/06/improving-speech-representations-and.html}), where google creates small personalized models for users (differently from our approaches).

% Customization? Can't really cite our HMM stuff...

% lots of lightweight fine-tuning stuff... LOTS (LORA, prefix-tuning, etc. etc.). Would be good to see if someone's already looked into how complementary the various approaches are...

% You can kind of elicit perf gains using prompt programming or wtv, but definitely helps to give more context. And gains are largest when you do actual training, and like you can't really attend to a whole repo (although you kind of can..., well can show that performer sort of failed..., anyway you can't with dense attention)

\section{Conclusion}
In this paper we explored different ways to customize a code generation model for a given codebase, with the goal of improving its performances on a target project. We described and analyzed four customization strategies and applied them on 20 different software projects for the task of generating unit test cases. 

Specifically, we considered the following strategies: (i) \textit{custom} fine-tuning, which allows all the model parameters to be tuned on the target project; (ii) \textit{L-EO} fine-tuning, a lightweight training which freezes most of the model's parameters, tuning only embedding and output layers; (iii) \textit{L-LDB} fine-tuning, a lightweight training which only tunes the last decoder block; (iv) \textit{prefix} tuning, which keeps language model parameters frozen, but optimizes a small project-specific vector (prefix).

In our extensive empirical evaluation we found that all the customization strategies lead to significant model's improvements on a target project, in terms of both intrinsic and task-specific metrics, with the custom models adapting to the coding style of the target project. While there is no clear winner among the customization strategies, each approach can provide specific benefits in particular deployment scenarios.

% \bibliographystyle{IEEEtran}

% \bibliographystyle{plainnat}
\bibliographystyle{ACM-Reference-Format}
\bibliography{main,references,emnlp2020}
\balance

\end{document}